\documentclass[prl,twocolumn,aps,groupedaddress,showpacs]{revtex4}
\usepackage{amsmath,amsfonts,amssymb}
\usepackage{wrapfig}
\usepackage{graphicx}
\usepackage{bbm}
\usepackage{graphics}

\def \beq {\begin{equation}}
\def \eeq {\end{equation}}
\def \ba {\begin{eqnarray}}
\def \ea {\end{eqnarray}}
\newcommand{\upp}{\hspace{-0.2 pt}\uparrow}
\newcommand{\downn}{\hspace{-0.2 pt}\downarrow}
\newcommand{\SV}{\hat{\vec{S}}}
\newcommand{\Sz}{\hat{S}_z}
\newcommand{\Jz}{\hat{J}_z}

\newcommand{\Az}{\hat{A}_z}

\newcommand{\Am}{\hat{A}_-}
\newcommand{\Ap}{\hat{A}_+}

\newcommand{\AV}{\hat{\vec{A}}}

\newcommand{\ketbrad}[1]{|#1\rangle\!\langle #1|}

\newcommand{\Ikz}{I_z^{(k)}}
\def\ket#1{\left| #1\right>}
\def\bra#1{\left< #1\right|}

\newcommand{\Up}{\uparrow}
\newcommand{\Dn}{\downarrow}
\newcommand{\cD}{{\cal D}}

\def\lsim{\mathrel{\rlap{\lower4pt\hbox{\hskip1pt$\sim$}}
    \raise1pt\hbox{$<$}}}                
\def\gsim{\mathrel{\rlap{\lower4pt\hbox{\hskip1pt$\sim$}}
    \raise1pt\hbox{$>$}}}                

\begin{document}

\title{Quantum information processing using localized ensembles of nuclear spins}
\author{J. M. Taylor$^1$, G. Giedke$^2$, H. Christ$^3$, B. Paredes$^3$, J. I. Cirac$^3$,
P. Zoller$^4$, M. D. Lukin$^1$, and A. Imamo\u{g}lu$^2$}
\affiliation{
$^1$ Department of Physics, Harvard University, Cambridge,
Massachusetts 02138, USA \\
$^2$ Institut f\"ur Quantenelektronik, ETH Z\"urich, Wolfgang-Pauli-Stra{\ss}e 16, CH-8093 Z\"urich, Switzerland \\
$^3$ Max-Planck-Institut f\"ur Quantenoptik, Hans-Kopfermann-Stra{\ss}e 1, Garching, D-85748, Germany\\
$^4$ Institut f\"ur Theoretische Physik, Universit\"at Innsbruck, Technikerstra{\ss}e 24, A-6020 Innsbruck, Austria}

\begin{abstract}
We describe a technique for quantum information processing based on localized ensembles of nuclear spins.  A qubit is identified as the presence or absence of a
collective excitation of a mesoscopic ensemble of nuclear spins surrounding
a single quantum dot.  All single and two-qubit operations
can be effected using hyperfine interactions and single-electron
spin rotations, hence the proposed scheme avoids gate errors arising
from entanglement between spin and orbital degrees of freedom.
Ultra-long coherence times of nuclear spins suggest that this scheme
could be particularly well suited for
applications where long lived memory is essential.
\end{abstract}
\maketitle

Nuclear spin degrees of freedom have attracted considerable
attention as potential carriers of quantum information due to
their exceptionally long coherence times. Early bulk NMR
work~\cite{CFH97} has substantially enriched our
understanding of the key features of quantum
computation~\cite{BCJ+99,ScCa99}. The fundamental difficulties in
scaling bulk NMR to a large number of qubits motivated efforts to
use single, individually addressable nuclear spins in
semiconductors as qubits~\cite{kane98}, where computation is
primarily mediated by the hyperfine interaction between electron
and nuclear spin. While possibly scalable, such a scheme is
limited by the fact that the electron wave-function is spread over
many lattice sites, reducing the strength of the hyperfine
interaction. In addition, two-qubit operations in
Ref.~\cite{kane98} rely upon exchange coupling, making them
susceptible to fast orbital decoherence mechanisms.

Recently, a method for robust storage of quantum information in
localized  ensembles of nuclear spins was
suggested~\cite{taylor03,taylor03b}, where it was shown that the
collective hyperfine coupling between nuclear and electron spin
degrees of freedom provides a controllable mechanism for coherent
storage and manipulation of quantum states.  These nuclear spin
ensembles correspond, for example, to the lattice nuclei in a
quantum dot. As a quantum memory, such nuclear ensembles are
robust with respect to variations in dot characteristics, rely
upon proven fabrication techniques, and provide high fidelity
storage without requiring a high-degree of nuclear spin
polarization.

In this Letter, we describe a technique to efficiently process
quantum information stored in localized nuclear spin ensembles.
Specifically, these ensembles enable a
robust, scalable implementation of quantum computation protocols,
unencumbered by the difficulties faced by single spin impurity or
bulk NMR approaches. 
The
fundamental interaction that allows for spin manipulation in our
scheme is hyperfine coupling: as a result, the orbital and spin
degrees of freedom remain unentangled throughout the two-qubit
gate operations, mitigating the effects of orbital decoherence on
gate fidelity. While collective enhancement of hyperfine
interaction allows for fast quantum gates, the ultra-long nuclear
spin coherence times render the scheme particularly attractive for
memory intensive quantum information processing tasks.

\begin{figure}
\includegraphics[width=2.6in]{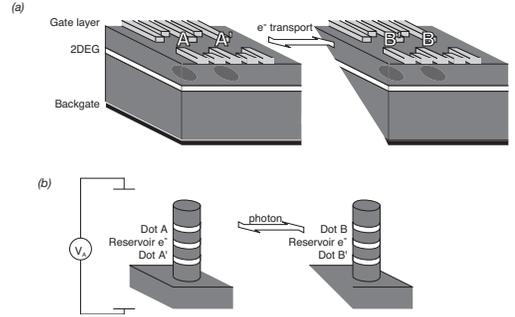}
\caption{ Quantum dot based  approaches for nuclear
ensemble-based computation.
(a)  Electrically defined (lateral)
quantum dots connected by ballistic transport {\em e.g.} through a series of quantum dots.  Nearby rf-SETs (not shown) would provide measurement.
(b) Nanowhiskers~\cite{BOS+02}
with several optically active dots.  Tasks between A
and A' are achieved via shuttling of an electron from
the reservoir dot via a potential difference, $V_A$.
Long distances tasks are performed through photon-based
entanglement generation between A' and B'.~\cite{CCG+99} \label{f:scheme}}
\end{figure}

The scheme proposed here can be realized using either
electrically~\cite{loss98} or
optically~\cite{IAB+99} manipulated quantum dots
that are defined as in Fig.~\ref{f:scheme}.  The nuclear ensemble in each dot is
prepared using polarized electron
spins~\cite{imamoglu03,taylor03b}. We illustrate  our technique by
considering the fully polarized case (when the nuclear state is
$\ket{0} = \ket{-I,\ldots,-I}$ for $N$ spin-$I$ nuclei), 
demonstrate how to perform single qubit and two-qubit gates, 
and consider sources of error.

The Hamiltonian describing a single electron interacting with the
nuclear spins in a quantum dot is  \cite{SKL03}
\begin{equation}\label{e:Horig}
H = \hbar B \gamma_S 
\hat S_z + \hbar B\sum_k \gamma_I^{(k)}\Ikz+\hbar a \SV \cdot
\AV.
\end{equation}
The first and second terms describe the coupling of the electron
and the nuclear spins to an external magnetic field, with
effective gyromagnetic  ratios $\gamma_{S},\gamma_I^{(k)}$.  The
last term is the electron coupling to collective nuclear degrees
of freedom, defined by $\hat A_{\pm,z} = \sum_k \alpha_k \hat
I_{\pm,z}^{(k)}$.  The $\alpha_k$'s are proportional  to the
probability density of the electron at the location of the
corresponding nuclei
and are normalized such that $\sum_k\alpha_k=N$; $a=A/N$ is the
average hyperfine interaction per nucleus.  Both $A$ and $N$
depend upon the specific material and dot construction; typical
numbers are $A \sim10-100$ ns$^{-1}$ and $N=10^4-10^6$ nuclei.
Excitations out of the fully polarized state form an orthonormal set
of collective nuclear spin states $\ket{m}\propto(\Ap)^m\ket{0}$,
where $m$ is the number of excitations.

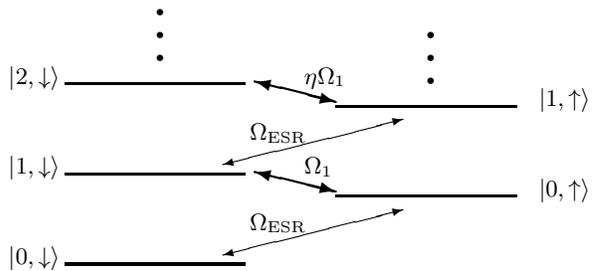
\begin{figure}
\unitlength3mm
\begin{picture}(30,10)
\thicklines
\put(3,1){\line(1,0){8}}
\put(0.5,1){$\ket{0,\Dn}$}
\put(3,5){\line(1,0){8}}
\put(0.5,5){$\ket{1,\Dn}$}
\put(3,9){\line(1,0){8}}
\put(0.5,9){$\ket{2,\Dn}$}
\put(15,4){\line(1,0){8}}
\put(24,4){$\ket{0,\Up}$}
\put(15,8){\line(1,0){8}}
\put(24,8){$\ket{1,\Up}$}
\thicklines
\put(11.4,5.1){\vector(4,-1){3.6}}
\put(11.4,5.1){\vector(-4,1){0}}
\put(13.6,5){$\Omega_1$}
\put(11.4,9.1){\vector(4,-1){3.6}}
\put(11.4,9.1){\vector(-4,1){0}}
\put(13.6,9){$\eta\Omega_1$}
\thinlines
\put(10,5.4){\vector(-4,-1){0}}
\put(10,5.4){\vector(4,1){8}}
\put(11.2,2.5){$\Omega_\mathrm{ESR}$}
\put(10,1.4){\vector(-4,-1){0}}
\put(10,1.4){\vector(4,1){8}}
\put(11.2,6.5){$\Omega_\mathrm{ESR}$}
\put(19,9){\tiny $\bullet$}
\put(19,10){\tiny $\bullet$}
\put(19,11){\tiny $\bullet$}
\put(7,10){\tiny $\bullet$}
\put(7,11){\tiny $\bullet$}
\put(7,12){\tiny $\bullet$}
\end{picture}
\caption{
  Level structure of the combined electron spin and nuclear dark state
  plus excited states.  For $N\rightarrow \infty$, the coupling
  between the two excitation manifold is stronger by $\eta \rightarrow
  \sqrt{2}$.
\label{f:singleQubit}
}
\end{figure}

When an electron is confined in the dot,  evolution over times
much shorter than $a^{-1}$ is restricted to 
subspaces spanned by $\{ \ket{m-1} \ket{\upp}, \ket{m}
\ket{\downn} \}$. Using Pauli matrices, the Hamiltonian for each
subspace with a fixed excitation number $m$ is $H^{(m)} = \hbar
\delta_m \sigma_z^{(m)} + \hbar \Omega_m \sigma_x^{(m)}$, with
\begin{subequations}
\ba
\Omega_m & = & a/2 \sqrt{\bra{m-1} \Am \Ap \ket{m-1} } ; \\
\delta_m & = & a/4 (\bra{m} \Az \ket{m} + \bra{m-1} \Az \ket{m-1})
 \\
& & \hspace{-6ex}+ [\gamma_S - (\bra{m}K_z\ket{m}-
\bra{m-1}K_z\ket{m-1})] B/2,\nonumber
\ea
\end{subequations}
where $K_z=\sum \gamma_I^{(k)}\Ikz$ is a sum over nuclear spin
operators, weighted by individual nuclear spin gyromagnetic
ratios. When Overhauser shift and Zeeman energy sum to zero
($|\delta_m| \ll \Omega_m$) and coherent flip-flop (Rabi)
oscillations occur at rate $\Omega_m$. The energy level structure
of the coupled electron-nuclear system is shown in
Fig.~\ref{f:singleQubit} along with the coupling strengths for $m
\le 2$: since $\Omega_{m+1} = \eta_{m} \Omega_m$ where $\eta_m =
\sqrt{m+1}[1-{\mathcal O}(m/N)]$, it is easy to note the analogy
with the celebrated Jaynes-Cummings (JC) model of quantum
optics~\cite{foot0}. We use the nonlinearity of such a JC-type
two-level system coupled to a nearly-bosonic mode to effect
elementary quantum gates.

Quantum information stored in the $m=0,1$ manifold can be mapped
reliably from nuclear states to electron spin and back
\cite{taylor03b} via a generalized rotation : \beq
R_{en}^{xy}(\pi/2,0) (\alpha \ket{0} + \beta \ket{1}) \ket{\downn}
= \ket{0} (\alpha \ket{\downn} + i \beta \ket{\upp}), \eeq where
\beq R_{en}^{xy}(\theta,\phi) = e^{i \phi \Sz} e^{-i \theta H/(
\hbar \Omega_1)} e^{-i \phi \Sz} . \eeq
The transfer of quantum
information from the nuclear ensemble to electron spin allows for
fast single qubit operations to be performed: after a
$\ket{\downn}$ electron is injected into the dot, the quantum
information is transferred to the electron spin,
and then the 
operation is performed on the electron. Finally, the quantum
information is mapped back to the nuclear ensemble. A $z$-axis
rotation can be accomplished by waiting in a static magnetic field
or through a laser-induced spin-dependent AC Stark
shift~\cite{imamoglu00}. $x$-axis rotations can be done via
ESR~\cite{loss98,VHB+02} (with $\Omega_\mathrm{ESR} \gtrsim 1$
ns$^{-1}$) or optical Raman spin flips through a virtual trion
state (with $\Omega_\mathrm{opt,ESR}
\gtrsim 10$ ns$^{-1}$)~\cite{IAB+99}.
Measurement of the ensemble nuclear spin state can be
implemented by mapping the quantum information to electron spin,
and carrying out an electron spin measurement either by
state selective ionization followed by charge measurement with a
rf-SET~\cite{hanson03b} or by detecting fluorescence in a
spin-dependent cycling transition~\cite{imamoglu00,GEL+03}.

To perform a two-qubit gate between quantum dots A and A', a
single electron can be used as to transfer quantum information between the dots:
the state of nuclear spin qubit A is mapped onto the electron,
which is then moved to A', where a controlled-phase (CP) gate between
the nuclear and the electronic qubit is applied using the nonlinearity
of the interaction. Following Ref.~\cite{ChCh00}, a two-qubit
CP-gate (up to single qubit gates) is given by
\beq
R^{xy}_{en}(\pi/4,0) R^{xy}_{en}(\pi/\eta,-\pi/2)R^{xy}_{en}(-\pi/4,0).
\eeq
In the computational basis,
this corresponds to
\begin{equation}
\left(
\begin{matrix}
1 & 0 & 0 & 0 \\
0 & e^{i \pi/\eta} & 0 & 0 \\
0 & 0 & e^{-i \pi/\eta} & 0 \\
0 & 0 & 0 & -1
\end{matrix}\right) \label{e:ent} .
\end{equation}
After this operation, the quantum information carried by the
electron spin is mapped back onto the nuclear spin ensemble A.
Applying a $z$-rotation of $-\pi/\eta$ to  qubit A and one of
$\pi/\eta$ to qubit A' yields a CP gate.

For distant dots, where electron spin  transport between ensembles
is difficult or impossible, it may be easier to generate
entanglement off-line, apply local purification protocols and then use
it to effect non-local gates, following \cite{CDKL00}.
Entanglement between distant electron spins can be generated
optically, using spin-flip optical Raman transitions
\cite{CCG+99}, or electronically, such as through adiabatic splitting
of a spin singlet in a double dot~\cite{taylor04epr}.
The electron-nuclear state mapping procedure can then ensure that the
distant nuclear spin ensembles are entangled.  Starting with the
entangled nuclear state
$\frac{1}{\sqrt{2}}\left(\ket{01}+\ket{10} \right)$ (between dots A'
and B',
cf.~Fig.~\ref{f:scheme}), we can implement a
CP gate on dots A and B (deterministically) by
performing local unitaries and measurements of AA' and BB' as follows:
(i) perform a
$\mathrm{CNOT}_{A\to A'}$ then measure A'. (ii) Perform a
CP gate $\ket{10}\to-\ket{10}$ at BB' followed by a
Hadamard gate and measurement at B'. (iii) Local phase flips
$\ket{0}\to -\ket{0}$ at A (B) if the measurement outcomes at
B'(A') were ``1'' complete the CP gate between A and
B~\cite{foot1}.

We next analyze various sources of  errors due to finite
polarization, inhomogeneity, nuclear spin dynamics, and electron
spin decoherence.  To understand the role of finite polarization,  the specific
cooling procedure must be examined.  When the nuclear ensemble
starts as a thermal mixture, cooling to dark states can be
achieved by coupling polarized electron spins to the nuclear
ensemble~\cite{imamoglu03}.  Regardless of the details of cooling,
the final density matrix will be a statistical mixture of dark
states $\ket{\cD (n,\beta)}$, where $\Am \ket{\cD (n,\beta)} = 0$.
$n$ is the number of spin excitations ($n=0$ is the fully
polarized state) and $\beta$ is the permutation group quantum
number~\cite{arecchi72}.  It was recently
shown~\cite{taylor03b} that dark states have the same symmetry
properties as the fully polarized state, and a manifold of excited
states can be defined from a given dark state, $\ket{m(n,\beta)}
\propto (\Ap)^m \ket{\cD (n,\beta)}$.  Hence the above
considerations for perfectly polarized nuclei map directly to the
case when the nuclear ensembles start in any given dark state, not
just the fully polarized state.

In practice, the cooled nuclear ensemble density matrix is a
mixture of different dark states, i.e. $\hat \rho = \sum_{n,\beta}
\rho_{n,\beta} \ketbrad{\cD (n,\beta)}$.  As each dark state has a
different $\delta_m(n,\beta),\Omega_m(n,\beta)$, interaction times
and applied magnetic fields can only produce
$R_{en}^{xy}(\theta,\phi)$ with the desired angle $\theta$ for
some fraction of the given mixture.  The inhomogeneous mixture
effects can be characterized by examining the subgroups of dark
states with different detunings, which lead to errors in Rabi
oscillations $p \simeq (\sigma_{\delta}/\Omega_1)^2$, where
$\sigma_{\delta}$ is the standard deviation of possible $\delta_1$
values over the distribution $\rho_{n,\beta}$.  In the homogeneous
case with spin-1/2 nuclei,
\begin{equation}
\sigma_{\delta}  \simeq  a \sqrt{N(1-P^2)}  \;\;. \end{equation}
Even at high ($P \sim 0.95$) polarizations, the effect of different detunings can be substantial ($p \sim 0.03$).  This provides the strongest limit to realization of
$R^{xy}_{en}$~\cite{foot2}.

The inhomogeneous nature of the hyperfine coupling leads to
further errors. In this case, the logical states of the system
(the $m$-excitation manifolds) are no longer eigenstates of $\Jz =
\Az+\Sz$ and of $\hat{A}^2$.
As a consequence, there is a nonzero probability that the system
moves out of the computational subspace during the gate operation.
We estimate these leakage errors using the techniques developed in
Refs.\cite{taylor03,taylor03b}, and find that the resulting gate
error $p_\mathrm{inhom}$ decreases with increasing number of
nuclei: for $N\sim 10^5$ at high polarizations ($P > 0.95$)
$p_\mathrm{inhom} \lesssim 10^{-3}$~\cite{foot3}. We note that a
similar error emerges due to the differences in Zeeman energy
associated to different nuclear species.  For materials like GaAs,
with gyromagnetic ratios varying greatly from species to species,
this limits the effectiveness of gate operations at high magnetic
field, resulting in the errors in the range of $10^{-3}$ as
indicated in Fig.3. Optical manipulation (e.g.\ tuning the system
into resonance via spin-dependent AC optical Stark shifts) may
mitigate this difficulty. Finally, in between gate operations the
errors associated with inhomogeneous evolution may be eliminated by
refocusing sequences of NMR pulses.

The nuclear spin diffusion due to dipole-dipole interactions with rate
$\gamma_{DD} \sim 60 {\rm ms}^{-1}$ leads to a decay of the coherences
which form the dark states~\cite{deng04}. Active corrections can be performed with
NMR pulse sequences that average the dipole-dipole Hamiltonian, such
as WaHuHa, improving the error rate to $\tau_{whh}^2
\gamma_{DD}^3$~\cite{whh, mehring76}.  After correction dark state
coherences could have lifetimes on the order of 0.1-1s for moderate
cycle times $\tau_{whh}$. Quadrupolar terms due to inhomogeneous
strain can cause additional differential phase evolution of different
nuclear spins.  However, this type of phase inhomogeneity leads to an
error second order in the interaction which is negligible ($\sim
10^{-7}$ per cycle of computation).

The errors of manipulation of single electron spins in quantum
dots via microwave or optical fields have been considered in
detail
elsewhere~\cite{loss98,VHB+02,IAB+99,GEL+03}, and only the relevant results are quoted here.  Electron
spin decoherence will most likely be limited by different
Overhauser shifts corresponding to different detunings, with error
going as $p \simeq(\sigma_{\delta}/\Omega_\mathrm{ESR})^2$; fast
ESR ($\sim 6\mathrm{ns}^{-1}$) will mitigate this effect; with
optical fields, even faster effective Rabi-oscillations are
predicted ($\Omega_\mathrm{opt,ESR}\sim50\mathrm{ns}^{-1}$) with
errors then limited by spontaneous emission to $\sim10^{-3}$. As
for measurement, the fundamental limit will be set by relaxation
of the electron spin, which has a time scale $\Gamma^{-1} \sim $
0.1-10ms, based on recent measurements of the spin relaxation
time~\cite{fujisawa01,hanson03b}.

Moving individual  electron spins over short distances incurs phase errors due to uncertainty in the
local nuclear field of the transport channel. The error induced by
randomly oriented nuclei during transit can be estimated by the
time-averaged hyperfine field the electron wavepacket encounters
over the whole process, $A / \sqrt{n L
h l}$, where $n$ is the density of lattice nuclei, $L$ is the
length of the channel, and $h l$ is the transverse area of the
channel.   For a 50nm wide, 10nm high channel, and a quantum dot
separation of $1\mu$m, the expected dephasing probability due to thermal
nuclei is $\simeq 5 \times 10^{-5}$ for a 3ns transfer time.

\begin{figure}
\includegraphics[width=2.6in]{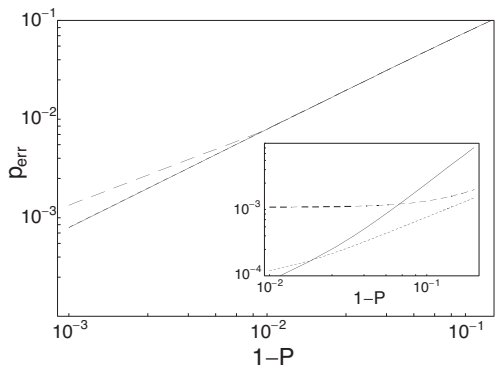}
\caption{ For a cooled dark state density matrix
  (c.f.~\cite{taylor03b}), expected error for a controlled-phase
  operation between electron spin and nuclear spin as a function of
  polarization (CdSe: solid, InAs: dotted, GaAs: dashed; in the main figure, CdSe and InAs are indistinguishable).  Inset shows effects of narrowing by a factor of 10
  for different materials.
\label{f:darkstategate}}
\end{figure}

We consider three materials with demonstrated electronic and optical
quantum dots (GaAs, $N=10^5$; InAs, $N=10^{4.6}$; CdSe, $N=10^4$).
The expected error of a two-qubit gate operation, defined as ${\rm
  Tr}[U_{\rm perfect} U_{\rm actual}^{\dag}]/4$, where $U_{\rm
  perfect}$ is given by Eqn.~\ref{e:ent}, is plotted in
Fig.~\ref{f:darkstategate}.  As the error is dominated by detuning
error, it is material independent, and is a few percent
at 95\% polarizations.

To combat these errors, a series of measurements may be made to
determine the effective detuning of the system better than the thermal
mixture limit $\sigma_{\delta}$.  This can be done {\em e.g.} by a
measurement made with the single electron in
the quantum dot, in direct analogy to single ion Ramsey interferometry in
atomic clocks~\cite{wineland}; the distribution is then narrowed, with
$\tilde{\sigma}_{\delta} \simeq \sigma_{\delta}/\sqrt{n}$, where $n$
is the number of measurements.  An improvement in the uncertainty
of the mixture's detuning of a factor of 10 yields
high fidelity operation, as shown in the inset to
Fig.~\ref{f:darkstategate}~\cite{foot5}.
Smaller dots produce better results at higher polarizations due to
their greater coupling strengths.
The limits for
GaAs are due to the large species inhomogeneity and the incommensurate
requirement of sufficient magnetic field to perform effective coherent
averaging of dipole-dipole interactions. Low species inhomogeneity
allows for higher magnetic field, faster WaHuHa-type correction
sequences, and fewer errors in the transfer operation. This would be
the case for quantum dots defined in nanotubes with isotopically
enhanced $^{13}$C or in CdSe quantum dots.  Materials with low
spin-orbit interaction will also reduce electron spin dephasing.

In conclusion, we have detailed a scheme for quantum information processing using
dynamically defined qubits composed of collective excitations of
nuclear spins in a quantum dot. For small dots with near-homogeneous
Zeeman splittings (e.g.\ CdSe, InAs) large but finite polarization
(95\%) is already sufficient to reach an error rate in two-qubit
operations of order a few percent.  Given the suppression of dipolar
diffusion in lithographically isolated structures (e.g.\  vertical
quantum dots, nanowhiskers, self-assembled quantum dots) these
polarizations may be within reach.
Moderate
polarizations~\cite{gammon01} have been already achieved, and techniques to
improve upon this have been considered~\cite{imamoglu03}.  Our
calculations indicate that the dominant source of error may be
mitigated through narrowing the mixture of dark states.

We gratefully acknowledge helpful conversations with
A. S{\o}rensen.  The work at Harvard was supported by ARO, NSF,
Alfred P. Sloan Foundation, and David and Lucile Packard
Foundation.  I. Cirac acknowledges DFG (SFB 631). The work at ETH
was supported by Nanoscience NCCR.


\begin{thebibliography}{30}
\bibitem{foot1}
Both imperfect entanglement and measurement errors lead only to phase-flip errors; thus, error correcting codes optimized for such errors may yield more favorable fault-tolerant thresholds.
\bibitem{foot2}
A similar analysis
shows that $\sigma_{\Omega} \simeq (a/\Omega_1) \sigma_{\delta} $
and is negligible in comparison to detuning-based errors.
\bibitem{foot3}
For a manifold based on a dark state $\ket{{\mathcal D}(n,\beta)}$ 
the value of $p_\mathrm{inhom}$ can be 
estimated, using Fig. 2 of Ref.~\cite{taylor03b}, 
as $(\chi_n^2 + \omega_n^2)/\Omega^2$.
\end{thebibliography}
\end{document}